\documentclass[12pt]{revtex4}
\begin{document}
\title{Comparison of different approaches of finding the positive definite metric in pseudo-Hermitian theories}


\author{ Ananya Ghatak \footnote{e-mail address: gananya04@gmail.com}}
\author{Bhabani Prasad Mandal \footnote{e-mail address:
 \ \ bhabani.mandal@gmail.com, \ \  bhabani@bhu.ac.in  }}


\affiliation { Department of Physics, \\
Banaras Hindu University, \\
Varanasi-221005, INDIA. \\
}

\begin{abstract}
To develop a unitary quantum theory with probabilistic description for pseudo-Hermitian systems one needs to consider 
the theories in a different Hilbert space endowed with a positive definite metric operator. There are different 
approaches to find such metric operators. We compare the different approaches of calculating positive definite metric 
operators in pseudo-Hermitian theories with the help of several explicit examples in non-relativistic as well as in 
relativistic situations. Exceptional points and spontaneous symmetry breaking are also discussed in these models.
\end{abstract}

\maketitle
\newpage
\section{Introduction}
Over the last decade complex extension of quantum mechanics has generated huge excitement \cite{int1,am}. It has been shown that certain categories of non-Hermitian operators can lead to fully consistent quantum theory with unitary time evolution if the associated Hilbert space is equipped with appropriate positive definite inner product rule. Development took place mainly in two directions. Bender et. al. have shown that certain non-Hermitian systems which are invariant under combined parity (P) and time reversal (T) transformation give rise to entire real spectrum \cite{int2}. Fully consistent quantum theory with unitary time evolution \cite{int4} for such systems is achieved by introducing CPT-inner product \cite {Cop1} for such systems. Operator C analogous to charge conjugation symmetry is associated with all PT-symmetric non-Hermitian systems. Many examples of physical systems in this category have been investigated in details \cite{pt3}-\cite{za1}. Consequences of spontaneous breakdown of PT-symmetry have recently been observed experimentally in optics \cite{opt1}-\cite{opt3}.  

Another class of non-Hermitian systems where the Hamiltonian H is related to its adjoint through a similarity transformation \cite{am}, 
\begin{equation}
H = S^{-1} H^{\dagger}S 
\label{h}
\end{equation}
have also been studied extensively \cite{spec1,spec2,ali2,rev1,ali,bp,bsg}. The adjoint $H^{\dagger}$ is defined with respect to the Hilbert space $\tilde{\cal H}$ equipped with the inner product 
\begin{equation}
\langle \phi\mid H\psi\rangle = \langle H^{\dagger}\phi\mid\psi\rangle.
\label{two}
\end{equation}
These pseudo-Hermitian systems are shown to reduce to PT-symmetric non-Hermitian systems when $S=P$. In such pseudo-Hermitian systems the energy eigenvalues are either real or appear in the complex conjugate pairs \cite{za1,spec1,ali2}. However the eigen vectors of H alone do not form a complete set of orthonormal functions and may not have positive definite norms with respect to the inner product in Eq. (\ref{two}). For any physical system the state vectors must have the positive definite norms to ensure the probabilistic description of quantum theory. Further it is not possible to have a unitary time evolution in such systems as the Hamiltonian is not self adjoint. To develop a consistent quantum theory (i.e., unitary quantum theory with probabilistic description) for such systems
one needs to define a new Hilbert space, ${\cal H}$ with a positive definite inner product. In all pseudo-Hermitian theories there is an additional operator $S$ [Eq. (\ref{h})] which allows us to define a new inner product, 
\begin{equation}
\left <\phi\mid\psi\right > _{S} = \left <\phi\mid\ S\mid\psi\right > = \left <S\phi\mid\psi\right >.
\label{w}
\end{equation}
The adjoint with respect to the above inner product is defined as, 
\begin{equation}
\langle H^{\ddagger}\phi\mid\psi\rangle_{S}=\left <\phi\mid\ H\psi\right >_{S} = \left <\phi\mid SH\psi\right >=\left <SS^{-1}H^{\dagger}S\phi\mid\psi\right >=\langle S^{-1}H^{\dagger}S\phi\mid\psi\rangle_{S}
\end{equation}
which implies that the adjoint with respect to this modified inner product in Eq. (\ref{w})  is ${H^{\ddagger} = S^{-1}H^{\dagger}S}$. Motivation for introducing such an inner product is clear from the fact that the Hamiltonian in pseudo-Hermitian theories [ as defined in Eq. (\ref{h})] are self adjoint, i.e., ${H^{\ddagger} = H}$ with respect to this new inner product. Thus one can have a unitary time evolution for pseudo-Hermitian theories with modified inner product
\footnotemark[1]\footnotetext[1]{We would like to point out that one does not require to construct positive definite inner product in the effective description of non-Hermitian theories. In particular, non-Hermitian theories which are considered in optics \cite{opt1}-\cite{opt3} are effective theories and hence standard metric is used instead of positive definite metric.}
\begin{equation}
\langle \phi (t)\mid\psi (t)\rangle_{S} = \langle e^{-iHt}\phi (0)\mid e^{-iHt}\psi (0)\rangle_{S} = \langle e^{+iH^{\ddagger}t}e^{-iHt}\phi (0)\mid\psi (0)\rangle_{S} = \langle \phi (0)\mid\psi (0)\rangle_ {S}.
\end{equation} 
However still we have a problem in developing probabilistic quantum theory with a pseudo-Hermitian theory, as even this modified inner product too may or may not lead to positive definite norms for the state vectors for the pseudo-Hermitian system. In order to overcome this problem one further needs to modify the Hilbert space with an inner product in which pseudo-Hermitian theories are self adjoint and the norms of the state vectors are positive definite. This can be achieved by constructing a new Hilbert space with a positive definite metric operator $\tilde{S}$. The construction of positive definite metric operator in pseudo-Hermitian theories includes that of CPT-operator \cite{Cop1} as a special case for PT-symmetric non-Hermitian theories.

There are several approaches to construct positive definite metric operator $\tilde{S}$ \cite{am,all}. The purpose of this work is to compare the different approaches with explicit calculation in different pseudo-Hermitian theories. We mainly focus on the spectral approach \cite{spec1,spec2} and on the approach developed by Das \cite{all}. By considering three explicit examples both in non-relativistic as well as relativistic quantum mechanics we verify that both the method lead to the same metric operator, modulo an overall constant in one of the examples. It is interesting to note that the spectral method is much simpler and elegant compare to the other approaches when the Hamiltonian is diagonalizable. However, spectral method fails when the Hamiltonian is non-diagonalizable. On the contrary Das's method as shown in \cite{all} can be used perturbatively to calculate the positive definite metric operator even when the non-Hermitian Hamiltonian is non-diagonalizable. We further discuss the occurrence of exceptional points and spontaneous symmetry breaking in these models.

The plan of the paper is as follows. We describe briefly the different approaches to construct the positive definite metric operator in Sec. II. In section 
III we compare the metric operators constructed in both the approaches for various models. Applicability of positive definite metric operators in non-Hermitian
quantum theories is discussed in Sec. IV. Section V is kept for concluding remarks.

\section{Different approaches to calculate metric operator}
The construction of the positive definite metric operator is the central problem in pseudo-Hermitian quantum mechanics. There are various methods of systematic construction of positive definite metric operator. In this section we briefly outline those methods. We restrict our discussion to two the most commonly used methods namely the spectral method \cite{am,spec1,spec2} and the method constructed by Das et al \cite{all}.
\subsection{Construction of positive definite metric operator $(q)$ by Das et al method}
It has been shown in Ref. \cite{all} that if there exists an arbitrary operator A, which commutes with the Hamiltonian, i.e. $\left [A,H \right] = 0$, then one can define a new inner product 
\begin{equation}
\left <\phi\mid\psi\right > _{q} = \left <\phi\mid\ q\mid\psi\right >
\end{equation}
where $q=SA$ for a S-pseudo-Hermitian theory. 
The adjoint with respect to this new inner-product is then defined as 
\begin {equation}
H^{\ddagger}=q^{-1}H^{\dagger}q=A^{-1}S^{-1}H^{\dagger}SA=A^{-1}HA=H.
\end{equation}
Since the Hamiltonian is now self-adjoint with respect to this newly defined inner product, time evolution would continue to be unitary.

The arbitrary operator A which commutes with Hamiltonian has been expressed in terms of projection operator in the energy space $P_{E}$ which projects $\psi_{E}$ to those states with positive definite norms \cite{all} as,
\begin{equation}
A = \sum_{E} c_{E} P_{E}
\end{equation}
with $P_{E}\mid\psi_{E}\rangle = \delta _{E E^\prime}\mid \psi_{E^\prime}\rangle$.

Now it is straight forward to see that A satisfies the relation $A\mid \psi_{E}\rangle = c_{E}\mid \psi_{E}\rangle.$
The inner product with respect to q operator can then be written as \cite{all},
\begin{equation}
\left <\psi _{E^\prime}\mid \psi_{E}\right >_{q}=\left <\psi _{E^\prime}\mid SA\mid \psi_{E}\right >=c_{E}\left <\psi _{E^\prime}\mid S\mid \psi_{E}\right >=c_{E}\left <\psi _{E^\prime}\mid \psi_{E}\right >_{S}=c_{E}e^{-iF(E)}\delta _{E E^\prime}.
\end{equation}
Therefore, if $c_{E} = e^{iF(E)},$ is chosen then
\begin{equation}
{\left <\psi _{E^\prime}\mid \psi_{E}\right > _{q} = \delta _{E E^\prime}}.
\end{equation} 
Thus one can have a positive definite inner product that allows for a probabilistic description with unitary time evolution. Now the positive definite metric q can further be written as,
\begin{equation}
q=SA=S\sum e^{iF(E)}P_{E}=Se^{iF(E)}\sum P_{E}=Se^{iF(E)}.
\end{equation}
To construct the metric operator q one needs to consider the eigen equations of the Hamiltonian H and $H^{\dagger}$ i.e., 
$H\mid\psi_{E}\rangle = E\mid\psi_{E}\rangle,$
$H^{\dagger}\mid\phi_{E}\rangle = E^{\ast}\mid\phi_{E}\rangle.$
An operator $\sigma_{E}$ is defined \cite{all} to generate the eigenstates of H with eigenvalue E, $\mid\psi_{E}\rangle = \sigma_{E}\mid\psi\rangle$, where $\mid\psi\rangle$ is a reference state. Further it has been shown by Das et. al \cite{all} that the eigenstates for $H^\dagger$ with energy $E^\star$ can be generated for a reference state $\mid\phi\rangle$ by the operator $\left (\sigma_{E^\star}^{\dagger}\right )^{-1}$ i.e.,
 $\mid\phi_{E}\rangle = \left (\sigma_{E^\star}^{\dagger}\right )^{-1}\mid\phi\rangle$. 
From (6) one can realize that the action of q can then be expressed as,
\begin{equation}
q\left(\sigma_{E}\mid\psi\rangle\right)= \left(\sigma_{E^\star}^{\dagger}\right)^{-1}\mid\phi\rangle.
\end{equation}
Which further leads to the explicit form of q \cite{all}, 
\begin{equation}
q = \sum_{E}\left(\sigma_{E^\star}^{\dagger}\right)^{-1}\ q_{0} \sigma_{E}^{-1}P_{E}.
\label{adq}
\end{equation}
The $q_0$ is given as $\left <\psi\mid\phi\right> =\left <\psi\mid\ q_{0}\mid\psi\right > = 1$.
This method looks complicated but does lead to positive definite inner product with unitary time evolution in non-Hermitian quantum mechanics. Moreover, this method can be used for  a perturbative construction of positive definite inner product even when H is non-diagonalizable. In that situation the pseudo-Hermitian Hamiltonian is written in the form 
\begin{equation}
H=H_0+\epsilon V(x), \ \ \ \mbox{$\epsilon$ is small}
\end{equation}
where $H_0$ is the part of the pseudo-Hermitian Hamiltonian $H$ which can be diagonalized. Unperturbed positive definite metric operator $q_0$ is calculated for the Hamiltonian $H_0$ then $\epsilon V(x)$ is treated perturbatively to calculate the correction in $q_0$. The positive definite metric operator is then calculated as an infinite power series in the coupling $\epsilon $. For detail calculation of $q$ we refer to the Das's work \cite{all}.
 
\subsection{Construction of positive definite metric operator $\eta$ by spectral method}
Spectral method introduced by Mostafazadeh \cite{am,spec1,spec2} is one of the simplest and straightforward methods to construct the positive definite $\eta$. This method is based on the spectral representation of the metric operator. For a Hamiltonian operator the eigenstates of a pseudo-Hermitian system do not satisfy orthonormality and completeness relations. Rather they follows bi-orthonormality condition,  
$\mid\phi_{n}\rangle\langle\psi_{m}\mid = \delta_{mn}$ and the completeness condition $\sum_{n}\mid\psi_{n}\rangle\langle\phi_{n}\mid = 1$ where, $H\mid\psi_{n}\rangle = E \mid\psi_{n}\rangle$ and $H^{\dagger}\mid\phi_{n}\rangle =E\mid\phi_{n}\rangle$.
In this spectral method the positive definite metric operator is calculated in a simple and elegant manner \cite{spec1,spec2} as, 
\begin{equation}
\eta = \sum_{n}\mid\phi_{n}\rangle\langle\phi_{n}\mid.
\label{et} 
\end{equation}
One can have a fully consistent quantum theory with pseudo-Hermitian Hamiltonian with the introduction of modified inner product with such a positive definite $\eta$. This method works fine as long as the Hamiltonian is diagonalizable.
\medskip
\section{Calculation of the positive inner product in different systems}
 In this section we consider explicit examples to calculate positive definite metric operator using both the methods discussed in previous section to make  a comparison between them. In the first example we consider a spin 1/2 system in the external magnetic field coupled to a simple Harmonic oscillator through pseudo-Hermitian interaction. Next we consider a general two level non-Hermitian system. Finally a pseudo-Hermitian scalar interaction in relativistic quantum mechanics is studied for comparison. It is possible to develop a fully consistent quantum theories with
 these non-Hermitian systems.

\subsection{System of spin 1/2 particle in an external magnetic field}
A system of spin 1/2 particles in the external magnetic field B coupled to an simple harmonic oscillator with frequency $\omega$ through the pseudo-Hermitian interaction \cite{bp} can be described by the Hamiltonian,
\begin{equation} 
H=\mu\sigma\cdot B+\hbar\omega a^{\dagger}a+\rho(\sigma_{+}a-\sigma_{-}a^{\dagger}).
\label{H}
\end{equation} 
Here $\sigma$'s are Pauli spin matrices, $\rho$ is some arbitrary real parameter, $\sigma_{\pm}=1/2(\sigma_{x}\pm i\sigma_{y})$ are spin projection operators. $a, a^{\dagger}$ are usual creation and annihilation operators for the simple Harmonic oscillator states. $a\mid n\rangle = \sqrt n\mid n-1\rangle, a^{\dagger}\mid n\rangle = \sqrt {n+1}\mid n+1\rangle$. $\mid n\rangle$ represents the eigenvectors for simple Harmonic oscillator. Without losing any essential feature of the system we can choose the external magnetic field along z-direction, the Hamiltonian will then change to,
\begin{equation}
H=\frac{\varepsilon}{2}\sigma_{z}+\hbar\omega a^{\dagger}a+\rho(\sigma_{+}a-\sigma_{-}a^{\dagger}).
\end{equation}  
It can be checked that this system is non-Hermitian $(H\not= H^{\dagger})$. However it is pseudo-Hermitian with respect to operator P and $\sigma_{z}$ i.e., $H^{\dagger}=PHP^{-1}$ and $H^{\dagger}=\sigma_{z}H\sigma_{z}^{-1}$.
One can check the ground state of the system in $\mid 0,-1/2\rangle$ with energy eigenvalue $=-\varepsilon/2$
\begin{equation}
H\mid 0,-1/2\rangle =-\varepsilon/2\mid 0,-1/2\rangle.
\end{equation}
Here we have applied the notation $\mid n,\frac{1}{2} m_{S}\rangle, n$ is the eigenvalue for the number operator $a^{\dagger}a$, i.e. $a^{\dagger}a\mid n,\frac{1}{2} m_{S}\rangle=n\mid n,\frac{1}{2} m_{S}\rangle$ and $m_{s}=\pm 1$ are the eigenvalues of the operator $ \sigma_{z}$, i.e. $\sigma_{z}\mid n,\frac{1}{2} m_{S}\rangle=m_{s}\mid n,\frac{1}{2} m_{S}\rangle$. The mechanism of $\sigma_{\pm}$ satisfying the following properties, \\
$ \sigma_{+}\mid n,1/2\rangle=0, \ \ \ \ \sigma_{+}\mid n,-1/2\rangle=\mid n,1/2\rangle,$ \\
$\sigma_{-}\mid n,-1/2\rangle=0, \ \ \sigma_{+}\mid n,1/2\rangle=\mid n,-1/2\rangle.$ \\
However $\mid 0,1/2\rangle$ is not a eigen state of this system but $\mid 0,1/2\rangle$ along with $\mid 1,-1/2\rangle$ creates an invariant subspace in the space of states as,
\begin{equation}
H\mid 1,-1/2\rangle=(\hbar\omega -\varepsilon /2)\mid 1,-1/2\rangle +\rho\mid 0,1/2\rangle.
\end{equation}
A general invariant subspace is consist of $\mid n,1/2\rangle$ and $\mid n+1,-1/2\rangle$ and the Hamiltonian matrix corresponds to this invariant subspace is,
\begin{equation}
H_{n+1} = \left(\begin {array}{clcr}
\epsilon /2+n\hbar\omega &  \rho\sqrt{n+1} \\
-\rho\sqrt{n+1}          &  -\epsilon /2+(n+1)\hbar\omega \\
\end{array} \right).
\end{equation}
The eigenvalues of the Hamiltonian matrix are given by \\  
\begin{equation}
\lambda^{\pm}_{n+1}=\frac{1}{2}\left[(2n+1)\hbar \omega\pm\sqrt{(\hbar\omega -\varepsilon)^{2}-4\rho^{2}(n+1)} \ \right]
\end{equation}
These eigenvalues are real provided $(\hbar\omega -\varepsilon)\geq 2\rho\sqrt{n+1}$
 and the normalized eigenstates in this regime are, 
\begin{eqnarray}
\mid\psi_{n+1}^{+}\rangle & = & \left(\begin {array}{clcr}
\sin {\theta_{n+1}}/2  \\
\cos {\theta_ {n+1}}/2  \\
\end{array} \right), \nonumber \\
\mid\psi_{n+1}^{-}\rangle & = & \left(\begin {array}{clcr}
\cos {\theta_{n+1}}/2  \\
\sin {\theta_ {n+1}}/2  \\
\end{array} \right)
\label{res}
\end{eqnarray}
where $\theta_{n+1}$ is defined as $(\hbar\omega -\varepsilon)\sin{\theta_ 
{n+1}}=2\rho\sqrt{n+1}$ to ensure the real eigenvalues. But when the strength of the non-Hermitian 
interaction is such that  $\rho>\frac{\hbar\omega -\varepsilon}{2\sqrt{n+1}}$, the eigenvalues of the $(n+1)$th
doublet become complex conjugate to each other and are written as,
\begin{equation}
E^{\pm }_n =\frac{1}{2}\left[(2n+1)\hbar \omega\pm i \sqrt{4\rho^{2}(n+1)-(\hbar\omega -\varepsilon)^{2}} \ \right]
\label{cev}
\end{equation}
The corresponding unnormalized eigenstates are,
\begin{eqnarray}
\mid\phi^+_{n+1}\rangle & = & \left(\begin {array}{clcr}
1  \\
\sin {\theta_{n+1}}+i \cos {\theta_ {n+1}}  \\
\end{array} \right), \nonumber \\
\mid\phi_{n+1}^-\rangle & = & \left(\begin {array}{clcr}
1  \\
\sin {\theta_{n+1}}-i \cos {\theta_ {n+1}}  \\
\end{array} \right)
\label{ces1}
\end{eqnarray}
The eigenstates in Eq.(\ref{res}) corresponding to the real eigenvalues do not form a complete set orthonormal eigenstates as expected in the case of 
pseudo-Hermitian systems. 
Rather these states form a complete sets of  bi-orthonormal states \cite{am}. 
However one can find positive definite metric operator for this system to have a consistent probabilistic quantum 
theory. Following the notations and methods in Sec 2.1, we consider $\mid\psi\rangle = \mid\psi_{n+1}^{-}\rangle$ and 
\begin{equation}
S = \left(\begin{array}{clcr}
1  & \ \ 0  \\
0 & -1 \\
\end{array} \right)
\end{equation}
where S is the similarity matrix for the present system which satisfies $H = S^{-1}H^{\dagger}S$,
to satisfy the condition $\langle\psi\mid\phi\rangle = 1$, $\mid\phi\rangle$ should be
\begin{equation}
\mid\phi\rangle = q_{0}\mid\psi\rangle.
\end{equation} 
So, ${\sigma_{E_{-}}=I, 
\sigma_{E_{+}}= \left(\begin{array}{clcr}
0 & 1 \\
1 & 0 \\
\end{array} \right)}$, and the projection operators for the wave functions with energies $E_{+}$ and $E_{+}$ can be calculated as $P_{E_{+}} = q_{0}\mid\psi_{+}\rangle\langle\psi_{+}\mid$ and $P_{E_{-}} = q_{0}\mid\psi_{-}\rangle\langle\psi_{-}\mid$. \\
Therefore we get the inner product using Eq.(\ref{adq}) as, 
$$q = (\sigma_{E_{+}}^{\dagger})^{-1}q_{0}(\sigma_{E_{+}})^{-1}P_{E_{+}} + (\sigma_{E_{-}}^{\dagger})^{-1}q_{0}(\sigma_{E_{-}})^{-1}P_{E_{-}}$$ 
This further can be written as
\begin{equation}
q = \left(\begin{array}{clcr}
1 & -\sin\theta_{n+1} \\
-\sin\theta_{n+1} & \ \ \ 1 \\
\end{array}\right).
\label{q}
\end{equation}
 Now we calculate the inner product by spectral method. For that we need to consider  
Hermitian conjugate of the Hamiltonian,
\begin{equation}
H_{n+1}^\dagger = \left(\begin {array}{clcr}
\epsilon /2+n\hbar\omega &  -\rho\sqrt{n+1} \\
\rho\sqrt{n+1}           &  -\epsilon /2+(n+1)\hbar\omega \\
\end{array} \right).
\end{equation}
The eigenvalues and eigenfunctions of $H_{n+1}^\dagger$ are
\begin{eqnarray}
E_{+} & = & 1/2\left[\hbar\omega\left(2n+1+\cos\theta_{n+1}\right)-\epsilon \cos\theta_{n+1}\right], \\ 
E_{-} & = & 1/2\left[\hbar\omega\left(2n+1-\cos\theta_{n+1}\right)+\epsilon \cos\theta_{n+1}\right],
\end{eqnarray}
\begin{eqnarray}
\mid\psi_{n+1}^{+}\rangle & = & \left(\begin {array}{clcr}
\cos {\theta_{n+1}}/2  \\
-\sin {\theta_ {n+1}}/2  \\
\end{array} \right), \nonumber \\
\mid\psi_{n+1}^{-}\rangle & = & \left(\begin {array}{clcr}
-\sin {\theta_{n+1}}/2  \\
\cos {\theta_ {n+1}}/2  \\
\end{array} \right).
\end{eqnarray}
We get the inner product using the Eq. (\ref{et}) for this case,
\begin{equation}
\eta = \mid \psi_{+}\rangle\langle\psi_{+}| \ + \mid\psi_{-}\rangle\langle\psi_{-}|
\end{equation} 
\begin{equation}
\eta = \left(\begin{array}{clcr}
1 & -\sin\theta_{n+1} \\
-\sin\theta_{n+1} & \ \ \ 1 \\
\end{array}\right)
\label{etaa}
\end{equation}
which is exactly same as the metric operator [ $q$ in Eq.(\ref{q})] obtained by Das's method.

An exceptional point occurs when $\theta_{n+1}=\frac{\pi}{2}$ as eigenvalues (Eq.(\ref{cev})) coalesce to real one and the
corresponding eigenfunctions (Eq.(\ref{ces1})) for the  $(n+1)$th doublet become,
\begin{eqnarray}
\mid\phi^+_{n+1}\rangle & = & \left(\begin {array}{clcr}
1  \\
\sin {\theta_{n+1}}  \\
\end{array} \right) = \mid\phi^-_{n+1}\rangle 
\end{eqnarray}
At the exceptional point 
det $\eta=0$ and the Hamiltonian can not be diagonalized. Alternatively the system passes from a broken 
symmetry phase to unbroken symmetry phase as the strength of the non-Hermitian interaction is reduced 
to $\rho<\left(\frac{\hbar\omega -\varepsilon}{2\sqrt{n+1}}\right)$.


\subsection{$2\times 2$ pseudo-Hermitian matrix Hamiltonian}
We consider a general ${2\times 2}$ PT-symmetric Hamiltonian \cite{all,prlc1} of the form,
\begin{equation}
H = \left(\begin{array}{clcr}
re^{\.{imath}\theta} & se^{\.{imath}\phi} \\
te^{-\.{imath}\phi} & re^{-\.{imath}\theta} \\
\end{array}\right)
\end{equation}
$r, s, t,$ are real parameters. The energy eigenvalues of these system are given as,
\begin{equation}
E_{\pm}=r \cos\theta \pm\sqrt{st-r^2\sin^2\theta}.
\end{equation}
In case of real eigenvalues we can define $E$ as $E_{\pm}=r \cos\theta\;\pm Q$, \ \ where $Q=\sqrt{st-r^2\sin^2\theta}=real$. 
The normalized wave functions for the systems are,
\begin{eqnarray}
\mid\psi_{E_{+}}\rangle &=& \frac{1}{\sqrt{s+t}}\left(\begin{array}{clcr}
(s/t)^{1/4}\sqrt{Q+\.{imath}r\sin\theta}e^{{\.{imath}\phi}/2} \\
(t/s)^{1/4}\sqrt{Q-\.{imath}r\sin\theta}e^{{-\.{imath}\phi}/2} \\
\end{array}\right), \nonumber \\
\mid\psi_{E_{-}}\rangle &=& \frac{\.{imath}}{\sqrt{s+t}}\left(\begin{array}{clcr}
(s/t)^{1/4}\sqrt{Q-\.{imath}r\sin\theta}e^{{\.{imath}\phi}/2} \\
-(t/s)^{1/4}\sqrt{Q+\.{imath}r\sin\theta}e^{{-\.{imath}\phi}/2} \\
\end{array}\right).
\end{eqnarray}
On the other hand when $st<r^2\sin^2\theta$, the eigenvalues become complex conjugate pair,
\begin{equation}
 E=r \cos\theta - i \tilde{Q} , \ \ \ \bar{E}=r \cos\theta + i\tilde{Q} 
 \label{cev2}
\end{equation}
where $\tilde{Q}=\sqrt{r^2\sin^2\theta-st}=real$. The corresponding eigenstates in this regime
are \cite{all},
\begin{eqnarray}
\mid\psi_{E}\rangle &=& \frac{-i}{\sqrt{(s+t)r\sin\theta+(s-t)\tilde{Q}}}\left(\begin{array}{clcr}
i\sqrt{s(r\sin\theta -\tilde{Q})}e^{{\.{imath}\phi}/2} \\
\sqrt{t(r\sin\theta +\tilde{Q})}e^{{-\.{imath}\phi}/2} \\
\end{array}\right), \nonumber \\
\mid\psi_{\bar{E}}\rangle &=& \frac{i}{\sqrt{(s+t)r\sin\theta+(s-t)\tilde{Q}}}\left(\begin{array}{clcr}
\sqrt{s(r\sin\theta +\tilde{Q})}e^{{\.{imath}\phi}/2} \\
-i \sqrt{t(r\sin\theta -\tilde{Q})}e^{{-\.{imath}\phi}/2} \\
\end{array}\right).
\label{ces}
\end{eqnarray}

Now we concentrate in the situation when energy eigenvalues are real. In this problem we identify
$S=\left(\begin{array}{clcr}
0 & e^{\.{imath}\phi} \\
e^{-\.{imath}\phi} & 0 \\
\end{array}\right)$
and $q_{0} = -\left(\frac{s+t}{2Q}\right)S$.
By calculating projection vectors $P_{E_{+}}$, $P_{E_{-}}$ and $\sigma_{E_{+}},\sigma_{E{-}}$ we find the positive definite inner product,
\begin{equation} 
q = \left(\frac{s+t}{2Q^{2}}\right)\left(\begin{array}{clcr}
t & -\.{imath}r\sin\theta e^{\.{imath}\phi} \\ 
\.{imath}r\sin\theta e^{-\.{imath}\phi} &  \ \ \ \ \ \ \ s \\
\end{array}\right).
\label{q2}
\end{equation}
In the spectral method we need the wavefunctions for $H^\dagger$ which are calculated as,
\begin{eqnarray}
\mid\psi_{E_{+}}\rangle &=& \frac{1}{\sqrt{s+t}}\left(\begin{array}{clcr}
(s/t)^{1/4}\sqrt{Q-\.{imath}r\sin\theta} \ e^{{\.{imath}\phi}/2} \\
(t/s)^{1/4}\sqrt{Q+\.{imath}r\sin\theta} \ e^{{-\.{imath}\phi}/2} \\
\end{array}\right), \nonumber \\
\mid\psi_{E_{-}}\rangle &=& \frac{\.{imath}}{\sqrt{s+t}}\left(\begin{array}{clcr}
(s/t)^{1/4}\sqrt{Q+\.{imath}r\sin\theta} \ e^{{\.{imath}\phi}/2} \\
-(t/s)^{1/4}\sqrt{Q-\.{imath}r\sin\theta} \ e^{{-\.{imath}\phi}/2} \\
\end{array}\right).
\end{eqnarray}
This leads to the expression for $\eta$ using equation (\ref{et}),
\begin{equation}
\eta = \frac{2}{t+s}\left(\begin{array}{clcr}
t & -\.{imath}r\sin\theta \ e^{\.{imath}\phi} \\ 
\.{imath}r\sin\theta \ e^{-\.{imath}\phi} &  \ \ \ \ \ \ \ \ \ s \\
\end{array}\right)
\label{et}
\end{equation}
which is same as $q$ [in Eq.(\ref{q2})] modulo a overall normalizing factor in this particular case.

For the specific values of r, s and t one encounters an exceptional point where $st=r^2 \sin^2\theta$ i.e, $\tilde{Q} =0$. At this
point two complex conjugate eigenvalues in Eq. (\ref{cev2}) coalesce to $E=r \cos\theta=\bar{E}$. The corresponding eigenstates in Eq. (\ref{ces}) 
also coalesce to one, modulo a overall normalizing factor,
$$
\mid\psi_{E}\rangle = \left(\begin{array}{clcr}
i\sqrt{s r\sin\theta}e^{{\.{imath}\phi}/2} \\
\sqrt{t r\sin\theta}e^{{-\.{imath}\phi}/2} \\
\end{array}\right) = \mid\psi_{\bar{E}}\rangle $$
At the exceptional point the system loose its completeness and the Hamiltonian can not be diagonalized as det $\eta =0$.
The condition $st>r^2 \sin^2\theta$ corresponds to the PT-unbroken phase of the system. On the other hand 
PT-symmetry is broken spontaneously when $st<r^2 \sin^2\theta$. We have fully consistent quantum theory in the modified Hilbert space endowed with 
the positive definite metric $\eta$ (Eq.(\ref{et})) when the system is in unbroken phase.

\subsection{Pseudo-Hermitian Scalar Interaction}
In all the previous examples we consider non-relativistic models. In this subsection we consider a relativistic model to arrive at the same conclusion. Let us consider a Dirac particle of mass ${m_{0}}$ subjected to a scalar pseudo Hermitian potential ${V_{s}}$, the dynamics can be described by the Hamiltonian \cite {bsg},
\begin{equation}
H=c\alpha\cdot p+\beta m_{0}c^{2}+V_{s}.
\end{equation}
For the sake of simplicity we take only one space dimension and choose he scalar pseudo-Hermitian potential as, ${V_{s}=v_{0}\left(\begin{array}{clcr}
0 &  1 \\
-1 & 0 \\
\end{array}\right)}$, $v_{0}$ is a real constant. Then we have,
\begin{equation}
H=\left(\begin{array}{clcr}
m_{0}c^{2} & cp_{x}+v_{0} \\
cp_{x}-v_{0} & -m_{0}c^{2} \\
\end{array}\right).
\end{equation}
We solve the Dirac equation to find energy eigenvalues, 
\begin{equation}
E_{\pm}=\pm\sqrt{\hbar^{2}c^{2}k_{x}^{2}+m_{0}^2c^{4}-v_{0}^{2}}=\pm E
\label{eng}
\end{equation}
and the eigenfunctions as,
\begin{eqnarray}
\mid\psi_{1}\rangle = \sqrt{\frac{E+m_{0}c^{2}}{2E}}\left(\begin{array}{clcr}
1 \\
\frac{cp_{x}-v_{0}}{E+m_{0}c^{2}} \\
\end{array}\right), \nonumber \\
\mid\psi_{2}\rangle = \sqrt{\frac{E+m_{0}c^{2}}{2E}}\left(\begin{array}{clcr}
-\frac{cp_{x}+v_{0}}{E+m_{0}c^{2}} \\
1 \\
\end{array}\right).
\end{eqnarray}
We identify ${S=\left(\begin{array}{clcr}
0 &  -1 \\
1 & \ 0 \\
\end{array}\right)}$ and ${q=\frac{E}{cp_{x}+v_{0}}I}$
and the ${\sigma_{E_{-}}=I, \sigma_{E_{+}}=\left(\begin{array}{clcr}
o & cp_{x}+v_{0} \\
cp_{x}-v_{0} & 0 \\
\end{array}\right)}$. So, we get the inner product as, 
\begin{equation}
q=\frac{E+m_{0}c^{2}}{2E}\left(\begin{array}{clcr}
1+\frac{(cp_{x}-v_{0})^{2}}{\left({E+m_{0}c^{2}}\right)^{2}} & \ \ \ \ \frac{2v_{0}}{E+m_{0}c^{2}} \\
\frac{2v_{0}}{E+m_{0}c^{2}} &   1+\frac{(cp_{x}+v_{0})^{2}}{\left({E+m_{0}c^{2}}\right)^{2}} \\
\end{array}\right).
\label{q1}
\end{equation}
On the other hand using spectral method, we can obtain $\eta$ by using Eq. (\ref{et})
where,
\begin{eqnarray}
\mid\psi_{1}\rangle = \sqrt{\frac{E+m_{0}c^{2}}{2E}}\left(\begin{array}{clcr}
1 \\
\frac{cp_{x}+v_{0}}{E+m_{0}c^{2}} \\
\end{array}\right), \nonumber \\
\mid\psi_{2}\rangle = \sqrt{\frac{E+m_{0}c^{2}}{2E}}\left(\begin{array}{clcr}
-\frac{cp_{x}-v_{0}}{E+m_{0}c^{2}} \\
1 \\
\end{array}\right)
\end{eqnarray}
are the eigenfunctions of 
\begin{equation}
H^{\dagger} =\left(\begin{array}{clcr}
m_{0}c^{2} & cp_{x}-v_{0} \\
cp_{x}+v_{0} & -m_{0}c^{2} \\
\end{array}\right),
\end{equation}
which leads to
\begin{equation}
\eta=\frac{E+m_{0}c^{2}}{2E}\left(\begin{array}{clcr}
1+\frac{(cp_{x}-v_{0})^{2}}{\left({E+m_{0}c^{2}}\right)^{2}} & \ \ \ \ \frac{2v_{0}}{E+m_{0}c^{2}} \\
\frac{2v_{0}}{E+m_{0}c^{2}} &   1+\frac{(cp_{x}+v_{0})^{2}}{\left({E+m_{0}c^{2}}\right)^{2}} \\
\end{array}\right),
\end{equation}
which is exactly same as $q$ obtained using the other method and given in the Eq. (\ref{q1}). From Eq. (\ref{eng})
it is clear that when $v_{0}^{2}>\hbar^{2}c^{2}k_{x}^{2}+m_{0}c^{4}$ the eigenvalues become imaginary and an exceptional point occurs at E=0, where even the
eigenfunctions become singular.

\section {Application of the Metric Operators}
PT-symmetric non-Hermitian system in the regime of unbroken symmetry and the pseudo-
Hermitian system can only lead to a fully consistent quantum theory if one finds the
positive definite metric operator associated with such systems. The state vectors form
a complete set of orthonormal functions and the norms of the eigenfunctions become
positive definite in the modified Hilbert space endowed with such positive definite metric
operator. The time evolution of the non-Hermitian system becomes unitary due to the
existence of such a positive definite metric operator. Therefore it is absolutely essential 
to construct a positive definite metric operator for any non-Hermitian system to make the 
theory physical.

In the first pseudo-Hermitian model we have considered a spin $1/2$ particle in the
external magnetic field B coupled to a simple harmonic oscillator which is commonly known 
as Jaynes-Cummings model and plays a very important role in quantum optics. It
describes in a simple way the interaction of photons with spin half particle. This model
has recently found an important application in quantum information theory where the
positive definite metric operator constructed for this model plays a crucial role.
The positive definite metric operator constructed for this model is extremely useful
to discriminate two non-orthogonal entangled quantum states \cite{ab}. If a particular quantum 
system is described by two states, $\mid\psi_{1}\rangle$ and $\mid\psi_{2}\rangle$ which are 
non-orthogonal but differ very slightly, i.e.
$$\langle\psi_{1}\mid\psi_{2}\rangle\not = 0, \ \ \ \ 
{\left|\langle\psi_{1}\mid\psi_{2}\rangle\right|}^{2}\cong 1-O(\epsilon^{2}), \epsilon\ll1 .$$
then it is not possible to determine the state of the system at a instant of time with a 
few measurements as $\mid\psi_{1}\rangle$ and $\mid\psi_{2}\rangle$ differ very slightly. 
This problem of quantum state discrimination is very important in quantum information theory 
\cite{qit}. It has been shown that these non-orthogonal states become orthogonal in a
modified Hilbert space endowed with a positive definite inner product associated with
the PT-symmetric non-Hermitian or pseudo-Hermitian systems. Alternatively these 
non-orthogonal states are allowed to evolve with a pseudo-Hermitian Hamiltonian in the usual 
Hilbert space to become orthogonal at some later time. However such a time evolution is 
obstructed by the possible existence of exceptional points in the non-Hermitian system 
\cite{ab} .

The model of spin $1/2$ particle which interacts with simple harmonic oscillator
in a pseudo-Hermitian manner is shown to be useful to discriminate two entangled
states of type \cite{ab},
\begin{eqnarray}
\mid\psi_{1}\rangle &=& \frac {1}{\sqrt{2}} \cos\frac{\theta}{2}\left [ \mid 0,1/2\rangle +\mid 1,-1/2\rangle \right ] + \frac {1}{\sqrt{2}} \ \sin\frac{\theta}{2} \left[\mid 0,-1/2\rangle +\mid 1,1/2\rangle\right ]; \nonumber \\
 \mid\psi_{2}\rangle &=& \frac {1}{\sqrt{2}} \cos\frac{\theta +2\epsilon}{2} \left[\mid 0,1/2\rangle +\mid 1,-1/2\rangle\right] + \frac {1}{\sqrt{2}} \sin\frac{\theta +2\epsilon}{2} \left[\mid 0,-1/2\rangle +\mid 1,1/2\rangle\right], \nonumber \\
\end{eqnarray}
where $\epsilon$  is a very small quantity and ${\left|\langle\psi_{1}\mid\psi_{2}\rangle\right|}^{2}\cong 1-\epsilon^2$.
The positive definite metric operator $q$ in Eq.(\ref{etaa}) for this model  
which was useful to make the theory fully consistent, discriminate
the above two non-orthogonal entangled states \cite{ab}. The positive definite metric operator associated with any non-Hermitian system 
can be used to discriminate states which differ slightly.


\section {Concluding Remarks}
Consistent quantum theory with pseudo-Hermitian Hamiltonian requires to find at least one positive definite metric operator. There are several approaches to 
construct positive definite metric operators. Considering several explicit examples both in non-relativistic as well as in relativistic  quantum mechanics we 
have verified that these approaches do lead to the same result. We have obtained the same expression for the positive definite metric operator in two of the 
examples. In the other example the positive definite metric operators obtained in different approaches differ only by a overall normalization constant. In our
 study we have paid attention to the two commonly used approaches namely spectral method and method by Das et al. We have also emphasized that it is easier to 
calculate the positive definite metric for diagonalizable non-Hermitian Hamiltonian using spectral method. On the other hand, spectral method fails for 
non-diagonalizable non-Hermitian systems, so one has to calculate positive definite metric perturbatively using Das's method. Further we discussed the 
possible existence of exceptional points in these models. The non-relativistic models 
become non-diagonalizable at the exceptional point. On the other hand the wavefunction become singular for the relativistic model at the exceptional point. 
Spontaneous symmetry breaking 
in all the three models have been discussed. Positive definite metrics associated with non-Hermitian theories play important role in quantum information
theory.


\begin{thebibliography}{99} 
 
\bibitem{int1} C.M. Bender, {\em Rept.Prog.Phys.} {\bf 70}, 947 (2007) and refs. therein.

\bibitem{am} A. Mostafazadeh, {\em Int. J. Geom. Math. Mod. Phys.} {\bf 7}, 1191 (2010) and references therein.
 
\bibitem{int2} C.M. Bender and S. Boettcher, {\em Phys. Rev. Lett.} {\bf 80}, 5243 (1998). 

\bibitem{int4} C.M. Bender, D.C. Brody and H. F. Jones, {\em Phys. Rev. D} {\bf 70}, 025001 (2004); Erratum-ibid. {\bf D 71}, 049901 (2005).


\bibitem{Cop1} C. M. Bender, K. Besseghir, H F Jones and X. Yin, {\em J. Phys. A} {\bf 42}, 355301 (2009); Carl M. Bender, Barnabas Tan, {\em J. Phys. A} {\bf 39} 1945 (2006); Carl M. Bender, Hugh F. Jones, {\em Phys.Lett.A} {\bf 328} 102 (2004).  


\bibitem{pt3} A. Khare and B. P. Mandal,{\em Phys. Lett. A} {\bf 272}, 53 (2000).
\bibitem{bag} B.Bagchi, C. Quesne, {\em Phys. Lett. A} {\bf 273}, 256 (2000).
\bibitem{pan} S. S. Ranjani, A. K. Kapoor and P. K. Panigrahi, {\em quant-ph} 0403054 (2004).
\bibitem{pan2} K. Abhinav, P. K. Panigrahi, {\em Annals Phys.} {\bf 326} 538 (2011). 
\bibitem{ss0} A. Mostafazadeh, {\em Phys.Rev. Lett.} {\bf 102}, 220402 (2009).
\bibitem{ss1} S. Longhi {\em Phys. Rev. B}  {\bf 80}, 165125 (2009).
\bibitem{ss2} B. F. Samsonov {\em J. Phys. A} {\bf 43}, 402006 (2010); B. F. Samsonov  {\em Math. J. Phys. A}: {\em Math. Gen.} {\bf 38}, L571 (2005).
\bibitem{pt31} M. Znojil and G. Levai, {\em Mod. Phys. Lett. A} {\bf 16}, 2273 (2001).
\bibitem{zno} M. Znojil, {\em Phys. Lett. A} {\bf 259}, 220 (1999); M. Znojil, {\em  J. Phys. A} {\bf 35}, 8793 (2002); ibid {\bf A 35}, 2341 (2002).
\bibitem{pij1} P. K. Ghosh and K. S. Gupta, {\em Phys. Lett. A} {\bf 323}, 29 (2004); P.K. Ghosh, {\em J. Phys.} {\bf A 38}, 7313 (2005).
\bibitem{zfr} A. Ghatak, J. A. Nathan, B. P. Mandal and Z. Ahmed, {\em J. Phys. A: Math. Theor.} {\bf 45} 465305(2012).
\bibitem{ab} A. Ghatak and B. P. Mandal, {\em J. Phys. A: Math. Theor.} {\bf 45} 355301(2012).
\bibitem{ab1} B. P. Mandal and A. Ghatak, {\em J. Phys. A: Math. Theor.} {\bf 45} 444022(2012).
\bibitem{ab2} B. P. Mandal, B. K. Mourya and R. K. Yadav (BHU) , {\em arXiv:1301.2387} (To appear in Phys. Lett. A).
\bibitem{bm} B. Basu-Mallick and B.P. Mandal, {\em Phys. Lett. A} {\bf 284}, 231 (2001).
\bibitem{sca} B. Basu-Mallick, T. Bhattacharyya  A. Kundu, and B. P. Mandal
{\em Czech. J. Phys } {\bf 54}, 5 (2004); B. Basu-Mallick, {\em Int. J. of Mod. Phys. B}  {\bf 16}, 1875 (2002).
\bibitem{sca1} B. Basu-Mallick, T. Bhattacharyya  and B. P. Mandal, {\em
 Mod. Phys.Lett. A} {\bf  20 }, 543 (2004).
\bibitem{new3} A. Khare and B. P. Mandal, {\em Spl issue of Pramana J of Physics} {\bf 73}, 387 (2009).
\bibitem{bn} Y. Brihaye and A. Ninimahazwe, {\em Int.J. Mod. Phys. A} {\bf 19}, 517 (2004); Y. Brihaye and A. Nininahazwe, {\em J.Phys.A} {\bf 40}, 13074 (2007). 
\bibitem{ft12} C. M. Bender, H.F. Jones and R. J. Rivers, {\em Phys. Lett. B}
{\bf 625}, 333 (2005). 
\bibitem{cavn} F. Cannata and A. Ventura, {\em J. Phys A: Math. Theor.} {\bf 41}, 
505305 (2008).
\bibitem{ben11} C.M. Bender, S. Boettcher and P.N. Meisinger, {\em J. Math. Phys.} {\bf 40}, 2210 (1999).  
\bibitem{new2} G. Levai, P. Siegl and M. Znojil, {\em Physics Letters A} {\bf 373}, 1921
(2009).
\bibitem{za1} Z. Ahmed, {\em  Phys. Lett. A} {\bf 324}, 152 (2004); ibid. A {\bf 294 }, 287 (2002).


\bibitem{opt1} Z. H. Musslimani, {\em Phys.Rev. Lett.} {\bf 100}, 030402 (2008).
\bibitem{opt2} A. Guo, G. J. Salamo, {\em Phys.Rev. Lett.} {\bf 103}, 093902 (2009).
\bibitem{lfng} L. Feng et. al {\em Science} {\bf 333}, 729 (2011); L. Feng et. al {\em Nature Materials}
{\bf 12}, 108 (2013). 
\bibitem{opt3} C. E. Rüter, K. G. Makris, R. El-Ganainy, D. N. Christodoulides, M. Segev, D. Kip, {\em Nature Physics} {\bf 6}, 192 (2010).


\bibitem{spec1} A. Mostafazadeh,  {\em J. Phys A: Math. and theor.} {\bf 36}, 7081 (2003).
\bibitem{spec2} A. Mostafazadeh and A. Batal, {\em J. Phys A: Math. and theor.} {\bf 37}, 11645 (2004).
\bibitem{ali2} A. Mostafazadeh , {\em J. Phys. A } {\bf 38}, (2005) 6657, {\em Erratum-ibid. A}{\bf 38}, 8185 (2005).
\bibitem{rev1} A. Mostafazadeh, {\em Ann. Phys} {\bf 309}, 1 (2004).
\bibitem{ali} A. Mostafazadeh, {\em J. Math Phys.} {\bf 43 } (2002) 205; {\bf 43} 2814 (2002); {\bf 43}, 3944 (2002).
\bibitem{bp} B. P. Mandal, {\em Mod. Phys. Lett. A} {\bf 20} 655 (2005).
\bibitem {bsg} B. P. Mandal, S. Gupta, {\em Mod. Phys. Lett. A} {\bf 25}, 1723 (2010).
 

\bibitem{all} A. Das, L. Greenwood, {\em J. Math. Phys.} {\bf 51}, 042103 (2010); A. Das, L. Greenwood, {\em Phys. Lett. B} {\bf 678 5}, 504 (2009).
\bibitem{prlc1} C.M. Bender, D. C. Brody and H. F. Jones {\em Phys. Rev. Lett.} {\bf 89}, 270401 (2002) ; Erratum-ibid. {\bf 92}, 119902 (2004).

\bibitem{qit} A. Chefles {\em Lect. Notes Phys.} {\bf 649}, 467 Springer (2004).
\bibitem{ben1} C. M. Bender, D. C. Brody, J. Caldeira, B. K. Meister, {\em arXiv} {\bf 1011.1871} (2010).





\end{thebibliography}
\end{document}